\documentclass[twocolumn, a4paper,fleqn,usenatbib]{aastex6}    
\usepackage[T1]{fontenc}
\usepackage{ae,aecompl}

\usepackage{graphicx}	
\usepackage{amsmath}	
\usepackage{amssymb}	
\usepackage{bm}
\usepackage{subfig}
\usepackage{times}
\usepackage{color}
\usepackage{soul}

\begin{document}
\title{An explosion is triggered by The late collapse of the compact Remnant from a Neutron Star Merger }

\author{Antonios Nathanail}
 
\affil{
Institut f\"ur Theoretische Physik, Goethe Universit\"at Frankfurt, \\
Max-von-Laue-Str.1, 60438 Frankfurt am Main, Germany; nathanail@th.physik.uni-frankfurt.de}


%
%
%
%







\begin{abstract}

It is known that a binary neutron star merger produces a hypermassive neutron star. 
The lifetime of this compact remnant  depends on the total mass and the equation of 
state. The collapse of this compact remnant to a black-hole-torus system is 
expected to give rise to a powerful jet and a short gamma-ray burst. 
Nevertheless, if the collapse is delayed half a second or so, the surrounding 
matter would be already accreted and/or expelled 
and hence no significant torus is formed. However, 
the collapse itself gives rise to a quasi-isotropic magnetized fireball.
This magnetic bomb  dissipates much of its energy due to magnetic 
re-connection and  produces the prompt emission. The energy range of such an explosion 
depends on the initial magnetic 
field strength and the amplification of the magnetic 
energy during merger. We briefly estimate the physical parameters 
at the time of collapse. We discuss the production of 
a quasi-isotropic magnetized fireball and its subsequent interaction 
with the ejected matter during merger, as the 
outcome of the coalescence of a binary neutron star system. We further 
suggest the radial stratification of the outflow, following the quasi-normal modes 
of the black hole.


\end{abstract}

\section{Introduction} 
\label{sec:intro}
The dawn of the multi-messenger era was marked by the observation of            
GW170817 and GRB170817A, with afterglow detection across the electromagnetic (EM) 
spectrum \citep{Abbott2017, Abbott2017b}. 
The simultaneous detection of  gravitational waves
(GW) and a short gamma-ray burst (GRB) made it clear 
that at least some short-GRBs are 
produced by binary neutron star mergers (BNS). 
It was a long standing conjecture that 
BNS are progenitors of short-duration GRBs \citep{Eichler89,Narayan92}.

One point of extreme interest is the overall low energetics of 
the prompt emission and the faint gamma-ray pulse, 
compared to the canonical short-GRB which pictures a highly 
relativistic outflow. After the merger of the BNS 
a hyppermassive neutron star (HMNS) is produced. The usual picture needs 
the collapse of this HMNS to produce a black hole 
and together with the surrounding torus to give rise to a 
relativistic jet expected.
However, the isotropic energy of GRB170817A observed to be $\sim 10^{46} {\rm  erg }$. 
The main arguments for this low flux 
detection is  picturing radiation from an off-axis jet, or radiation coming 
from a cocoon produced by the jet  while drilling through the ejected 
matter\footnote{Numerical simulations of BNS systems, 
with a total mass above $\sim 3.2 M_{\odot}$, have shown that after 
the merger, a magnetic-jet structure from the black-hole-torus system is produced 
\citep{Rezzolla:2011, Ruiz2016, Kawamura2016}. 
This is the starting point for  
 a jet  from a BNS merger \citep{Aloy:2005} and for off-axis radiation
 \citep{Lazzati2017a, Lazzati2017b, Kathirgamaraju2018} and coccon 
 emission \citep{Gottlieb2018}.}
 
 The first to consider the interaction of  a jet and the BNS ejecta 
 was \cite{Nagakura2014} and \cite{Murguia-Berthier2014}. They incorporated 
 a density profile, inspired from BNS simulations, mimicking the BNS ejecta 
 and the subsequent drilling of the jet through the ejecta. These can provide 
 jet conditions that allow the outflow to break out from the ejecta or not.

In what follows, we  describe the outcome of a
 BNS merger, where the merger remnant 
has a lifetime  of some seconds (even a fraction of a second). 
At the time of the collapse 
the surrounding matter consists 
a negligible torus.
Following this path, no jet is expected to form.
We  discuss  the collapse of the merger remnant 
 as the central engine of a magnetic explosion 
that drives a short-GRB. Most of 
the amplified, during merger, magnetic energy 
is released within a millisecond during collapse. This  
forms a magnetised fireball. 
\begin{figure*}
  \begin{center}
    \includegraphics[width=0.95\textwidth]{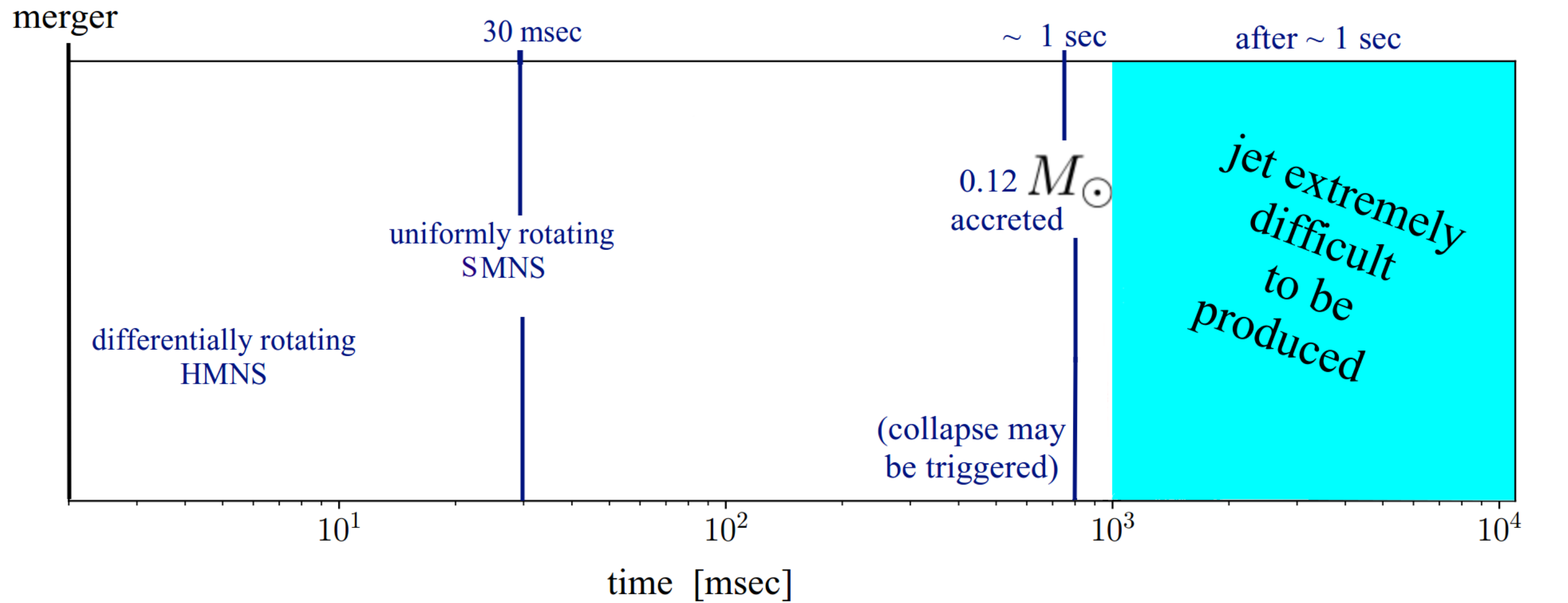}
  \end{center}
  \caption{We show the lifetime of the SMNS, if collapse is 
  triggered after $\sim 1{\rm  s } $, then it is extremely difficult to produce a jet.}
    \label{fig:life}
\end{figure*}

 In section \ref{sec:HMNS} we  estimate the physical parameters 
 of the merger remnant and the surrounding torus at  $\sim 1{\rm  s }$, 
 then present the main 
 features of the collapse  and discuss the energetics of such explosions. 
In section \ref{sec:model} we discuss 
the imprint of this  explosion.
In section \ref{sec:conc} we  conclude.

%
\section{The outcome of the merger and the collapse of the SMNS} 
\label{sec:HMNS}
%
%
The outcome of the coalescence of a BNS 
 can have different paths, which in turn would have different 
observational imprints. The difference in the collapse time of the HMNS, 
or a stable supra-massive neutron star (SMNS)\footnote{A SMNS is 
a NS above the TOV maximal mass which can still be supported by rigid rotation.}
 configuration could explain the short-GRB 
phenomenology \citep{Rezzolla2014b, Ciolfi2014, Usov1992, Metzger2008}.

Electromagnetic outputs of merging objects have been considered 
in the literature. Especially after the collapse of the SMNS, 
where the strong magnetic field  interaction  
with the black hole  that lasts a considerable amount of time, 
can shape  the afterglow of a short-GRB \citep{Lyutikov:2011c}.
In this paper we  focus on the possibility that 
the SMNS produced during merger,  collapses in a few seconds  after 
merger. The important point is not exactly the lifetime of the SMNS itself, 
but the parameters of the torus surrounding it.
The matter of the 
collapsing star itself  quickly 
hides behind an event horizon, but the magnetic energy 
stored in the near zone magnetosphere of the SMNS is released 
in a millisecond scale producing an outward magnetic shock. 
The magnetic 
energy of the SMNS is expected to be more than $10^{50}$ {\rm erg }, due to 
magnetic field amplification during merger.
Another meaningful point to note, is the 
difference from the collapse of an isolated NS 
which is expected to give rise to a short radio transient
\citep{Falcke2013}.
The major differences are two, on one hand  the enhanced magnetic energy 
due to amplification during merger, and on the other hand  the fact, that even
if no torus is formed, the SMNS is not isolated. It rather lives inside 
a dust of left-over  matter.
It has been argued that 
this pollution may not allow the magnetar model to function
\citep{Murguia-Berthier2014}. 
It is exactly this baryon 
pollution that contribute to the production of a fireball. \cite{Lehner2011} 
proposed that the collapse of the SMNS can produce an 
electromagnetic transient comparable 
in energy with a jet from a black-hole-torus when the SMNS collapses, 
and could contribute to the production of 
a short-GRB. 
They performed simulations for the gravitational collapse of an
isolated magnetized neutron star and showed the instantaneous (millisecond scale) 
dissipation of the magnetic energy. They further discussed in detail 
how this picture would evolve in the presence of an accretion disc, 
which is the case for the remnant of a BNS merger. They compared the 
electromagnetic luminosity prior to merger, with the post merger phase 
were the SMNS has a large but not coherent magentic field. Then, 
they estimated the 
electromagnetic transient produced from the collapse of the merger remnant 
and compared with the subsequent jet produced 
by the trapped magnetic flux between the newly formed black hole 
and the accretion disc \citep{Lehner2011}.

Following this thinking and these estimations, we  show that if the lifetime 
of the SMNS is longer than a second or so, then the limited amount of mass 
of the torus cannot result in a jet, but the collapse 
of the SMNS alone will power the short-GRB. 

 The total mass 
 (where in the case of GW170817 was 
 $\sim  2.74 \, M_{\odot}$ \cite{Abbott2017}) 
 and the EOS play the major role for the path 
that the merger product will take. 
Magnetohydrodynamic turbulence 
resulting in a effective viscosity 
can account for angular momentum transport which would leave 
a uniformly rotating SMNS after $\sim 30\, msec$.
Estimating this viscous angular-momentum transport yields 
\begin{equation}
\begin{aligned}
t_{{\rm MRI}} \sim \frac{R^2}{\nu}\sim 30 \, msec \left( \frac{R}{15\,{\rm  km }}
\right)^2  \left(\frac{H}{7.5 \,{\rm  km }}\right) \\ \times
\left( \frac{\alpha}{0.01}\right)^{-1}\left(\frac{c_s}{c/3}\right)^{-1} \,,
\end{aligned}
\end{equation}
where $\nu = \alpha c_s H\,$ is the viscous parameter, 
$c_s$ the speed of sound, $H$ the typical vertical scale height and the 
$\alpha$ parameter \citep{Shakura1973}. 
Thus, the SMNS is uniformly rotating in less 
than $30\, msec$ \citep{Fujibayashi2017}\footnote{The ADM mass of 
the SMNS in this study is $\sim 2.65 \, M_{\odot}$, and thus is relevant 
to the BNS that produced GW170817.}, whereas the outer part 
constitute  a quasi-keplerian disk \citep{Hanauske2016}.

The next important step is to estimate the evolution of the SMNS and 
the surrounding torus. 
If the HMNS survives for $30\, msec$, then differential rotation is lost 
and the maximum mass can be estimated for a uniformly rotating star. 
However, it is important to take into account the extra thermal 
pressure which can be $2\%$ of the total, 
especially  in the first stages after merger.
The cooling of the SMNS due to neutrino diffusion can be estimated: 
\begin{equation}
\begin{aligned}
t_{\nu}^{diff} \simeq 3 \,  \tau_{\nu} \,  \frac{\Delta R}{c} 
\sim 0.8 \,{\rm  s } \left( \frac{\tau_{\nu}}{10^3}\right) 
\left(\frac{\Delta R}{20 \,{\rm  km }}\right) \,,
\end{aligned}
\end{equation}
where  $\Delta R$ is the local density scale height and 
$\tau_{\nu}$ is the optical depth, with a mean value of $10^3$ 
for neutrinos of $10 - 100 \, MeV$ \citep{Dessart2009}.
This means that in around $1{\rm  s }$ there is a $2\%$ 
reduction of pressure, which could trigger the collapse of 
the SMNS. However, a study from \cite{Kaplan2013} showed that models 
with more thermal support are less compact, which make them more 
stable in  a longer timescale.

At this time the structure of the surrounding torus has also 
changed significantly. The viscous accretion timescale estimated
for the torus: 
\begin{equation}
\begin{aligned}
t_{accr} \simeq \, 1 \,{\rm  s } \left( \frac{R_T}{50\,{\rm  km }}
\right)^2 
\left(\frac{H_T}{25 \,{\rm  km }}\right)^{-1} \\ \times
\left( \frac{\alpha}{0.01}\right)^{-1}\left(\frac{c_s}{0.1c}\right)^{-1} \,
\end{aligned}
\end{equation}
where $R_T$ is the radius of the torus and $H_T$ is the 
typical vertical scale height of the torus. Then, the mass accretion rate 
onto the SMNS yields 
\begin{equation}
\begin{aligned}
\dot{M}_{SMNS} \simeq \frac{M_T}{t_{accr}} 
\sim 0.2 \,  M_{\odot} s^{-1} \left( \frac{\alpha}{0.01}\right)
\left( \frac{M_T}{0.2 M_{\odot}}\right) \\
\times
\left( \frac{R_T}{50\,{\rm  km }}
\right)^{-2} \left(\frac{H_T}{25 \,{\rm  km }}\right)\,,
\end{aligned}
\end{equation}
where $M_T$ is the mass of the torus. However, this accretion 
rate is not stationary, since the mass of the torus 
decreases in time and the torus expands. The radius of the 
torus can reach $140\,{\rm  km }$ in $1{\rm  s }$. 
Nevertheless, it can account for a mass accretion 
of $\sim 0.12 M_{\odot}$ in  $1{\rm  s }$ \citep{Fujibayashi2017}.

The previous estimate is another 
indication that collapse could be triggered at around $1{\rm  s }$ after merger,
due to accretion\footnote{As an example, for the LS220 EOS an equal 
mass  BNS of total mass
$M= 2.87\, M_{\odot}$, collapses straight after merger, whereas for 
a total mass $M= 2.67\, M_{\odot}$ does not collapse for 
some tens of milliseconds \citep{Bovard2017}.  }.
Furthermore, the effective viscosity inside 
the torus  results in the expansion of the torus. The  
density in the vicinity of the SMNS, at the time of collapse,
is  extremely important. As we  show below, this is the 
parameter that  designates the outcome of 
the collapse, a black-hole-torus magnetic jet or 
an induced magnetic explosion. The  
density of the torus at $1{\rm  s }$ is estimated by its left 
over mass and its expanded radius:
\begin{equation}
\begin{aligned}
\rho_T \simeq \frac{M_T}{2 H_T \pi R_T^2} 
\sim 9.2\times 10^{9} g/cm^3 \left( \frac{M_T}{0.08 M_{\odot}}\right)
\\ \times 
\left( \frac{R_T}{140\,{\rm  km }}\right)^{-2}  
\left(\frac{H_T}{70\,{\rm  km }}\right)^{-1}\,,
\end{aligned}
\end{equation}
where all quantities are for the expanded torus at $1{\rm  s }$ 
after merger. We should point out here, that if the collapse 
of the SMNS occurs even later, then  
there is the possibility that no debris disk is formed at all
\citep{Margalit2015}.

Before discussing about the production of a jet or a magnetic 
explosion, we should first have an estimate of the mean magnetic field 
strength of the SMNS. It is expected that during 
the merger process small scale turbulence can amplify the magnetic field in values 
higher than $10^{16} G$ \citep{Rasio99,Zrake2013b, Giacomazzo:2014b}.  
In such extreme conditions the magnetic energy  can     
reach as high as   $\sim 10^{51}{\rm erg }$ \cite{Kiuchi2015}. 

We  check what  happens, if the SMNS collapses 
after $1{\rm  s }$. One of the most important points, 
is the condition for the establishment of 
a magnetic jet. A stable configuration to be built by the 
black-hole-torus system needs that, at least, 
 the torus pressure can balance the magnetic pressure.
Following the above discussion, we assume that the mean magnetic field  
of the SMNS is $B \simeq 3\times 10^{16} \, G$. This yields:

\begin{equation}
\begin{aligned}
\frac{B^2_{SMNS}}{8 \pi}  \simeq 3.5 \times 10^{31}\, dyn/cm^2
\left( \frac{B_{SMNS}}{3\times  10^{16} \, G }\right)^2 \\
\gg 9.2 \times 10^{29} \, dyn/cm^2  \left( \frac{\rho_T}{9.2\times
  10^{9} \, g/cm^3 }\right)
   \simeq \rho_T \, c^2\,.
\end{aligned}
\end{equation}
This is estimated at around $1{\rm  s }$ after merger. It is evident 
that at later times, when the torus has expanded  more, 
the density  decreases and the establishment of a magnetic jet 
 becomes more difficult. A  lower estimate we get from  
 the accretion rate at  
$1{\rm  s }$, which is  $\sim 0.02 \, M_{\odot} s^{-1}$ 
as reported in \cite{Fujibayashi2017}.  This yields:
$$B^2_{SMNS}/8 \pi \gg 2.6 \times 10^{28} \, dyn/cm^2  
 \sim 
\dot{M}c/4\pi r_{BH}^2.$$
The above discussion is summarized in figure \ref{fig:life}.
So far, we have estimated that if the collapse is triggered around or 
after 
$\sim 1{\rm  s }$ after merger, the magnetic energy 
 of the SMNS is released and induce 
 a powerfull explosion of $E_{exp} \sim 10^{51} {\rm erg}$, 
contrary to a magnetic jet expected. The energetics for this magnetic bomb 
follows from the properties of the remnant (SMNS) itself. 
 The absence of a jet was discussed by \cite{Salafia2018}, where instead 
a flare produced during the 
 magnetic field amplification, gave rise to a relativistic isotropic fireball, 
which could explain 
the prompt emission of GRB170817A (\cite{Salafia2017}, see also \cite{Tong2017}).
Whereas, what we discussed here comes after merger and the absence 
of a jet is due to the physical conditions that come around or after $\sim 1{\rm  s }$.  
%
\begin{figure}
  \begin{center}
    \includegraphics[width=0.8\columnwidth]{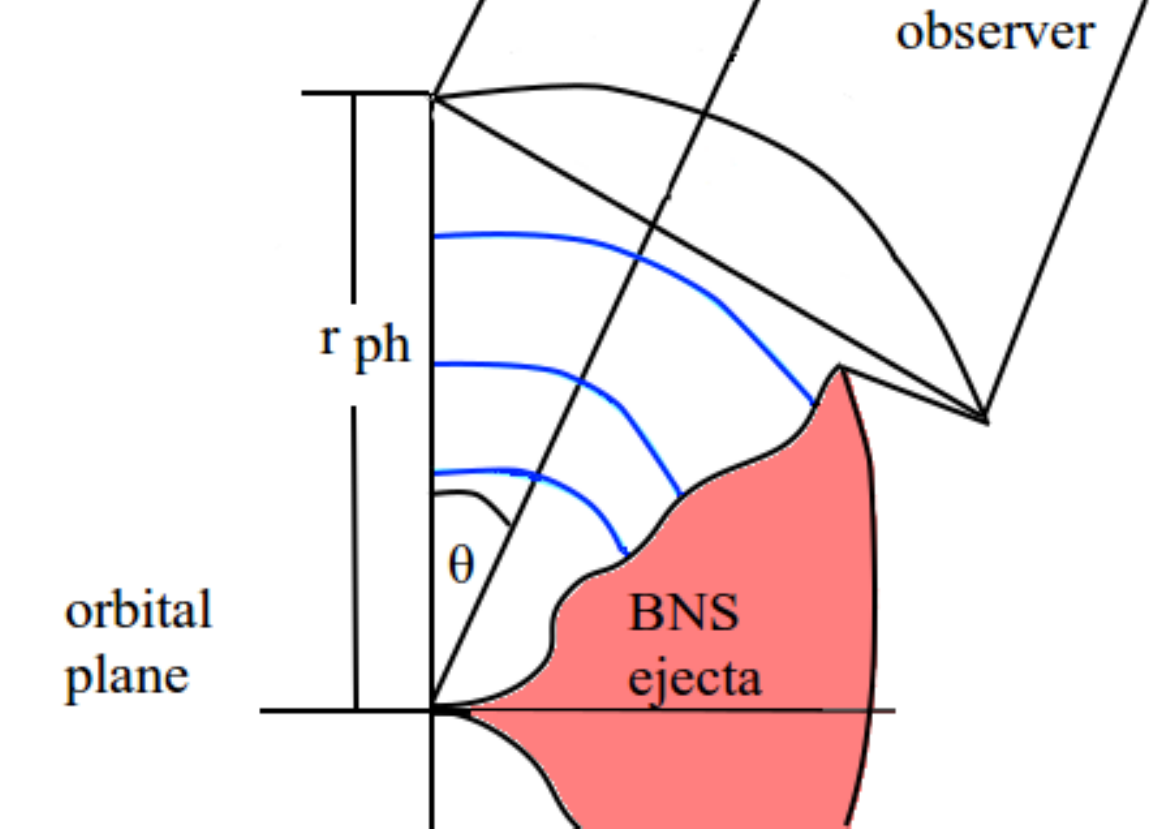}
  \end{center}
  \caption{In higher latitudes, the low density region allow 
  the quasi-spherical fireball to escape and reach the 
  transparency radius, $r_{ph}$. With blue color, we depict the slower cells of 
  the fireball produced by the "ring-down" of the black hole.}
    \label{fig:plot}
\end{figure}

The amount 
of baryons in this magnetic bomb is subject to 
the leftover matter exterior the SMNS 
and the neutrino-driven wind. Very quickly this magnetic bomb  
catches up the  relatively slow moving ejecta and crash on to them. 
This  matter, expelled dynamically 
or secularly,   is responsible for  the appearance of a 
kilonova \citep{Abbott2017c} and moves with velocities 
$u \simeq 0.1-0.2 c$ \citep{Metzger:2010, Bovard2017}. 

The matter ejected during merger depends on the EOS, nevertheless its averaged 
distribution on a sphere  far from the merger point shows that above the equator the 
overall ejected matter may be three orders of magnitude less than what was ejected 
in the equatorial plane (roughly $\sim 10^{-6}M_{\odot}$, \cite{Bovard2017}). The 
magnetic bomb  finds it easier to pass through the ejecta from higher latitudes. 
This means that it may have a wide angle jet structure, or rather that it is quasi-
isotropic with a big opening angle from the orbital axis. 

We have to stress out that what was discussed in this section is not at all any 
kind of a proof that the collapse of the merger remnant came at $\sim 1 \,{\rm  s }$ 
after merger. The major point here was that if the collapse comes after 
$\sim 1 \,{\rm  s }$, then it is extremely difficult to produce a magnetic jet.

\section{The imprint of the magnetic bomb} 
\label{sec:model}

\begin{figure}
  \centering
  \includegraphics[width=0.477\textwidth]{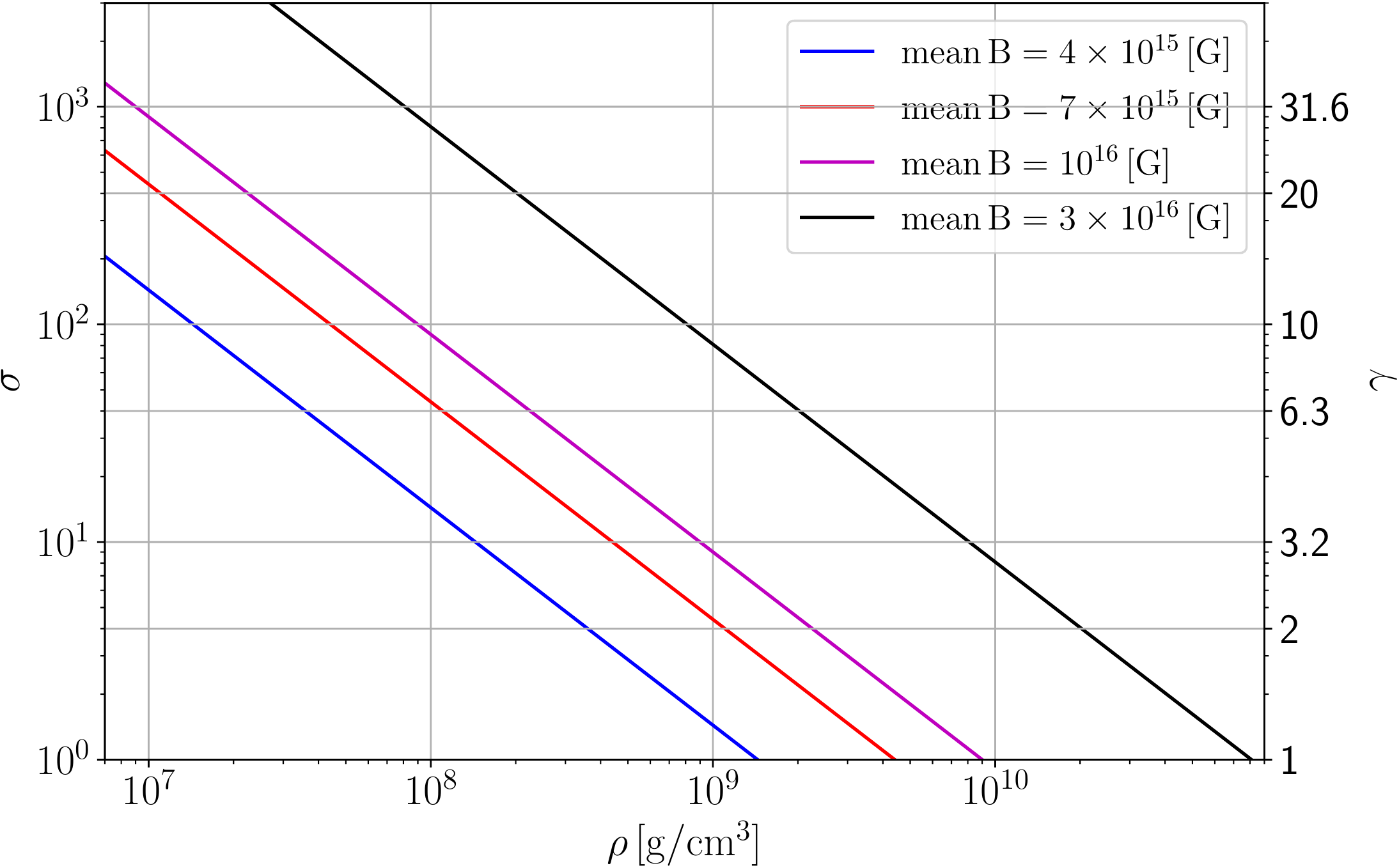}
  \caption{The magnetization parameter $\sigma$ (\textit{left column}) and the  
  Lorentz factor $\gamma$ (\textit{right column}), that corresponds to such
  magnetization in the downstream region, for 
  the relevant values of the density $\rho$, for different values of the 
  mean magnetic field $B$.}
  \label{fig:sigma}
\end{figure}

The formation of a magnetised fireball that crashes on the ejecta 
is established, if the collapse of the SMNS happens at  
$\sim 1 \, {\rm s }$.
In this section we   discuss the evolution of this magnetic bomb.
We show that after the explosion a relativistic outflow 
is produced. We further discuss  how this outflow 
can push matter from the polar region and surpass the 
BNS ejecta.
The amplified magnetic field in the SMNS is favored in 
the  toroidal direction and could produce a magnetically driven 
plasma gun \citep{Contopoulos1995}. At the time of collapse 
all this energy is released in a millisecond scale.
This explosion induces a shock to the surrounding matter. 
The shock front is perpendicular to the radial direction and parallel 
to the predominantly toroidal magnetic field. We assume that the 
upstream pressure is small, since the 
SMNS has collapsed when the magnetic explosion is produced.
Analysis of the jump 
conditions across the shock can yield useful estimations for 
the parameters of the post-shock region and have been widely 
used for astrophysical purposes (e.g. \cite{Kennel1984}).
We define the magnetization parameter as:
\begin{equation}
\begin{aligned}
\sigma = \frac{B^2}{4\pi \rho c^2}\,
\end{aligned}
\end{equation}
then the corresponding downstream Lorentz factor is 
\begin{equation}
\begin{aligned}
\gamma \simeq \sqrt{\sigma}
\end{aligned}
\end{equation}
In figure \ref{fig:sigma} we plot the magnetization 
parameter for the relevant values of the magnetic field and the density, 
where the density values can be viewed as an angular distribution of the 
matter from the equator till the polar region, 
where density falls significantly.
It is evident, that for a polar density of $10^7 {\rm g /cm^3}$ 
(as reported in \cite{Fujibayashi2017}) an outflow with a Lorentz 
factor of $10$ is formed. 

The next question to address, is whether this outflow can break out 
form the ejecta. In the case of a relativistic flow interacting with 
an external medium,  the Sedov length signals the radius where deceleration 
begins. 
The Sedov length is generally defined as the radius where the 
energy of the swept up matter: 
$E = m_p c^2 \int_0^l4\pi \rho r^2 dr$, equals the energy of 
the explosion.
The polar regions with less density are favored as a pathway for 
the explosion. As mentioned before, the ejected mass from the polar 
regions is on the the order of $\sim 10^{-6}M_{\odot}$, \citep{Bovard2017}.
Thus, the rest mass energy of the matter in the polar region is 
$E_{mpolar} \sim 10^{48} {\rm erg}$. Since the energy of the explosion 
is $E_{exp} \sim 10^{51} {\rm erg}$, the deceleration radius 
is reached further out in the inter-stellar medium.
and the outflow can successfully surpass the 
BNS ejecta from the polar region.

 The prompt gamma-ray emission consists of 
 a non-thermal pulse produced by 
accelerated particles through magnetic reconnection that will 
follow the collapse and the release of the magentic energy of the SMNS. 
During the expansion of the fireball, further dissipation of the magnetic energy
can contribute more to this. 
This will be followed (or accompanied) by a thermal high energy pulse 
produced almost with the fireball itself due to the pressure and the high 
temperatures in the post shock region. 
When the fireball reaches the 
photosphere ($r_{ph}  \sim 10^{12} {\rm cm}$), 
it is already further beyond the slow moving ejecta (at a 
similar time the ejecta is almost an order of magnitude 
closer to the source) and thus photons 
will not find any other obstacle(fig. \ref{fig:plot}). 

The duration of the high-energy pulse, 
which in the case of GRB170817A was $\sim 2 {\rm s }$ \citep{Goldstein2017},
The duration of the emission from the spherical cap is 
$dt\sim h/c = r_{ph}\times (1-cos(\theta))/c \simeq 2 r_{ph} \theta_f^2/c$.
For a mildly relativistic outflow with $\Gamma > 1/\theta_f$, 
the relevant timescale of the pulse is $dt\sim r_{ph} / 2 c \Gamma^2 $ 
\citep{Piran:2004ba}, 
which for a range of $\Gamma \simeq 7 - 10$, the duration 
is $dt\sim 1 - 2 {\rm s }$. 
 We should also mention here that even in the canonical 
short-GRB picture the $\sim 2 {\rm s }$ duration may be a problem if the disk 
is not massive enough and is accreted in less than a second.


Another point that we want to  touch is a   way  
of radially stratifying the fireball. It is known that the production 
of a black hole  is followed by the 
"ringing down" of the black hole \citep{Kokkotas99a_url}. 
The first burst from the collapse is followed by further pulses with 
almost an order of magnitude less intensity and a millisecond duration 
\citep{Lehner2011, Dionysopoulou:2012pp,Most2017}. 
Every pulse will be 
accompanied  by a production of a slower fireball, due to the reduced magnetic 
pressure injected by the pulse. The quasi-normal modes of the black hole 
are exponentially decaying, but can contribute to the radial stratification 
of the fireball. 
The fireball
may also form a coccoon-like structure at the intersection with the high-density 
ejecta and may have  a similar 
EM signature with a coccoon produced by a jet \citep{Gottlieb2018}.

 There is no 
need to speculate more about the radiation imprint of such fireball, 
since this should follow from a complete study of such a process.
Nevertheless, we should add that several observational facts that 
followed GW170817 and GRB170817A can accommodate the production of such 
wide-angle mildly relativistic outflow. The rising X-ray emission 
($\sim 3$ days after the event) and the 
radio  observations ($\sim 20$ days) can be explained with a 
quasi-isotropic outflow, and in the case we discussed
 the kinetic energy can be up to  the order of $10^{50} {\rm erg }$
 \citep{Alexander2017, Margutti2017}. Furthermore, it was stated 
 that the observed radio data have no direct indication of the presence 
 of a jet, and they can be explained by a wide-angle mildly 
 relativistic outflow \citep{Mooley2018}. Observations 
 (radio, optical and X-ray) cannot rule out 
 one or the other choice \citep{Margutti2018}.





 

%
\section{Conclusions} 
\label{sec:conc}
The aim of this note is to present and discuss the possibility 
that the outcome of a BNS merger follows 
a path that produces a quasi-spherical fireball behind the BNS ejecta 
rather than a jet. The main outcome of a merger 
discussed in the literature is the production of 
a black hole together with a surrounding torus or 
a stable magnetar, both of them are used in order to explain 
short-GRBs. In light of the new observations, we propose that it 
is equally possible to follow a third path, namely that the 
compact remnant, produced after merger, collapses and no significant 
torus is produced. This has as a consequence that all the 
magnetic energy of the SMNS will be released in a 
millisecond scale. The important point here, is that this 
energy can be as much as $10^{50}erg$, due to 
magnetic field amplification 
during merger. Thus, energetically it could in principle 
power a short-GRB \citep{Fong2015}. 


\section*{Acknowledgements}

It is a pleasure to thank D. Giannios, C. Fromm, O. Porth and L. Rezzolla
for useful discussions.   
The author is supported by an 
Alexander von Humboldt Fellowship.

\section*{}
\bibliographystyle{mnras}

\label{lastpage}
\end{document}